\documentclass[conference]{IEEEtran}
\IEEEoverridecommandlockouts
\usepackage[T1]{fontenc}
\usepackage{textcomp}

\usepackage{newtxtext,newtxmath}
\usepackage{cite}
\usepackage{algorithmic}
\usepackage{graphicx}
\usepackage{xcolor}
\usepackage{booktabs}
\usepackage{enumitem}
\usepackage{tabularray}
\definecolor{LightBlue}{RGB}{212, 230, 246} % Example
\newcommand{\cmark}{{\checkmark}}
\newcommand{\xmark}{{\textit{X}}}

\usepackage{hyperref}

\usepackage{amsmath}
\usepackage{mathtools}
\usepackage{caption}

\usepackage[ruled,vlined]{algorithm2e}

\usepackage{microtype}
\usepackage{graphicx}
\usepackage{subfigure}
\usepackage{booktabs} % for professional tables

\definecolor{mypink}{HTML}{FB2E99}

\author{
\IEEEauthorblockN{
Yancheng Zhang\IEEEauthorrefmark{2},
Mengxin Zheng\IEEEauthorrefmark{2},
Xun Chen\IEEEauthorrefmark{3} \\
Jingtong Hu\IEEEauthorrefmark{4},
Weidong Shi\IEEEauthorrefmark{8},
Lei Ju\IEEEauthorrefmark{9},
Yan Solihin\IEEEauthorrefmark{2},
Qian Lou\IEEEauthorrefmark{2}\IEEEauthorrefmark{1}
}

\IEEEauthorblockA{
\IEEEauthorrefmark{2}University of Central Florida,
\IEEEauthorrefmark{3}Samsung Research America, \\
\IEEEauthorrefmark{4}University of Pittsburgh,
\IEEEauthorrefmark{8}University of Houston,
\IEEEauthorrefmark{9}Shandong University \\}
\thanks{\IEEEauthorrefmark{1}Corresponding author: Qian Lou (qian.lou@ucf.edu)}
}

\def\BibTeX{{\rm B\kern-.05em{\sc i\kern-.025em b}\kern-.08em
    T\kern-.1667em\lower.7ex\hbox{E}\kern-.125emX}}

\begin{document}

\title{zkVC: Fast Zero-Knowledge Proof for Private and Verifiable Computing
}
\maketitle

\begin{abstract}
In the context of cloud computing, services are held on cloud servers, where the clients send their data to the server and obtain the results returned by server. However, the computation, data and results are prone to tampering due to the vulnerabilities on the server side. Thus, verifying the integrity of computation is important in the client-server setting. The cryptographic method known as Zero-Knowledge Proof (ZKP) is renowned for facilitating private and verifiable computing. ZKP allows the client to validate that the results from the server are computed correctly without violating the privacy of the server's intellectual property. Zero-Knowledge Succinct Non-Interactive Argument of Knowledge (zkSNARKs), in particular, has been widely applied in various applications like blockchain and verifiable machine learning. Despite their popularity, existing zkSNARKs approaches remain highly computationally intensive. For instance, even basic operations like matrix multiplication require an extensive number of constraints, resulting in significant overhead. In addressing this challenge, we introduce \textit{zkVC}, which optimizes the ZKP computation for matrix multiplication, enabling rapid proof generation on the server side and efficient verification on the client side. zkVC integrates optimized ZKP modules, such as Constraint-reduced Polynomial Circuit (CRPC) and Prefix-Sum Query (PSQ), collectively yielding a more than 12-fold increase in proof speed over prior methods. The code is available at \href{https://github.com/UCF-Lou-Lab-PET/zkformer}{\textcolor{mypink}{https://github.com/UCF-Lou-Lab-PET/zkformer}}.
\end{abstract}

% https://anonymous.4open.science/r/zkformer-5E69/

\begin{IEEEkeywords}
Private and Verifiable Computing, Zero-Knowledge Proof, Machine Learning
\end{IEEEkeywords}

\section{Introduction}
\label{introduction}
%Deep neural networks in machine learning have gained significant interest and achieved notable success in recent years. 

% what is zkp  + the use cases (for private and verifiable computing, e.g., linear projection (matrix multiplication) and ML)
% zkp is slow, (e.g., how slow for SoTA matrix multiplication and ML, overhead over regular computation)
% reasons and challenges for efficient ZKP, or what are ignored by prior works 
% our methods and why we can solve the challenges
% experiments on micro-benchmark matrix multiplication, and applications like ML, Transformer;

Zero-knowledge proof (ZKP)~\cite{parno2016pinocchio, groth16} is a cryptographic primitive that enables a prover to convince a verifier of the correctness of a computation without revealing the prover's secret input. ZKP ensures that a proof passes verification only if the computation was performed correctly, guaranteeing both verifiability and privacy. By providing strong guarantees on the integrity of computations while preserving the privacy of the server's input, ZKP has found broad applications in domains where computational integrity is critical, such as blockchain and verifiable machine learning.

\begin{figure}[h]
% \vspace{-0.4in}
    \centering
    \includegraphics[width=0.9\linewidth]{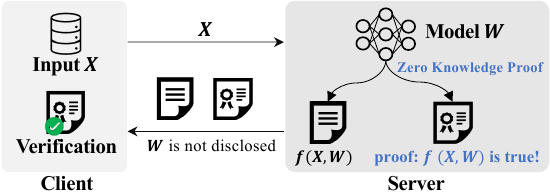}
     %\vspace{-0.1in}
    \caption{Example of use case in verifiable and private neural network inference based on Zero-knowledge Proof (ZKP).  
    }
    \label{fig:nizk}
 %\vspace{-0.2in}
\end{figure}

In Figure \ref{fig:nizk}, we illustrate one use case of ZKP in verifiable machine learning. ZKP allows the owner of a proprietary machine learning model, \textit{prover}, to prove to the users, \textit{verifier}, that predictions have been accurately computed by the pre-trained model, without compromising the model's confidentiality. As depicted, the client-side verifier first sends input data $X$ to the server, where the server's prover performs the neural network inference $f(X,W)$ to output a prediction. ZKP is then used to generate a proof for this inference, ensuring $f(X, W)$'s validity without revealing the model weights, enabling the client to verify the proof.

While promising, ZKP often leads to large computation overhead in practice, especially when proving matrix multiplication. Proving a single matrix multiplication of dimension $[49, 320]\times[320,512]$ on a 16-Core CPU with a commonly used ZKP scheme~\cite{groth16} known for efficiency, can take up to $\sim 3$ minutes. Matrix multiplication is a fundamental operation in numerous applications, including data analysis and machine learning. However, the inefficiencies in current ZKP frameworks make it impractical to directly scale them for real-world scenarios involving massive matrix multiplications. For instance, Transformer-based models~\cite{vaswani2017attention, dosovitskiy2020vit}, widely used in machine learning, rely heavily on extensive matrix multiplications. Proving the correctness of a single inference using a ViT-Base model~\cite{dosovitskiy2020vit} on the ImageNet dataset~\cite{deng2009imagenet} results in prohibitively high computational costs, rendering such approaches infeasible.

%\sim10Ω hours runtime. 

A series of works have explored how to scale up ZKP-based verifiable computations for real-world applications~\cite{lee2020vcnn, liu2021zkcnn, kang2022scalingzkml, weng2023pvcnn}, where many of them are built upon zk-SNARKs. zk-SNARKs is a branch of general ZKP,  which features short proof sizes, fast verification, and non-interactivity and is suitable for cloud computing. The main challenge in reducing the overhead of zk-SNARKs is to reduce the number of constraints. Recent works~\cite{lee2020vcnn, weng2023pvcnn} reduce proving complexity by expressing convolution as polynomial multiplication within a polynomial quadratic arithmetic program (QAP). However, their optimization highly depends on the convolution's unique feature, which cannot directly extend to general matrix multiplication. The efficient construction of zk-SNARKs for matrix multiplication remains an open problem.

Proving the correctness of matrix multiplication begins by transforming it into an arithmetic circuit, which is then represented as a QAP. The efficiency of zk-SNARKs relies heavily on the complexity of the QAP, determined by two key factors: (1) the number of constraints, dictated by the multiplication gates in the arithmetic circuit, which also defines the degree of the constraint system, and (2) the number of variables, corresponding to the length of a full assignment to the circuit. Previous work~\cite{lee2020vcnn} proposed reducing the number of constraints in convolution operations by introducing additional variables, referred to as dummy terms. However, directly applying this approach to matrix multiplication significantly increases the variable count, reducing overall efficiency. Notably, similar to the construction in vCNN~\cite{lee2020vcnn}, our verification process remains succinct and is largely independent of the complexity of the original computation.

\begin{figure}
    \captionsetup{skip=1.2pt}
    \centering
    \includegraphics[width=1\linewidth]{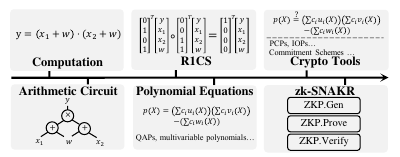} 
     %\vspace{-0.25in}
    \caption{The workflow of zero-knowledge proof systems.}
    \label{fig:zkp}
     \vspace{-0.25in}
\end{figure}

\noindent \textbf{Our Contributions.} This paper introduces \textit{zkVC}, an efficient zk-SNARK construction tailored for general operations such as matrix multiplication and their applications in machine learning, including attention-based Transformers. We propose \textbf{Constraint-Reduced Polynomial Circuits (CRPC)} to minimize the constraints required for matrix multiplication in ZKP. By transforming matrix multiplication into polynomial multiplication represented in a quadratic arithmetic program (QAP), CRPC reduces the number of constraints from $O(n^3)$ to $O(n)$. Additionally, we introduce \textbf{Prefix-Sum Query (PSQ)}, a technique that reduces the number of variables by optimizing the circuit for product accumulation in matrix multiplication. Experimental results demonstrate that the combination of CRPC and PSQ achieves a $12\times$ improvement in proving time for matrix multiplication compared to prior methods. Furthermore, we apply \textit{zkVC} to verifiable Transformer inference, achieving over a 15$\times$ runtime reduction on ViT models compared to baselines without our optimizations.

%Our techniques enable zkFormer to be efficiently proven less than $1\ hour$ and verified with only $94.2\ ms$, all while achieving $80\%$ top-1 accuracy and $95\%$ top-5 accuracy on ImageNet-1K. 

\section{Background and Motivation}

\noindent \textbf{ZKP Systems.}
ZKP is a cryptographic protocol enabling a prover $\mathcal{P}$ to assure a verifier $\mathcal{V}$ of a statement's truth without disclosing anything beyond its validity. In verifiable neural networks, it lets a model owner $\mathcal{P}$ confirm the accuracy of a neural network inference to a client $\mathcal{V}$. Figure \ref{fig:zkp} outlines a ZKP system's process. To prove a computation like $y = (x_1+w)\cdot(x_2+w)$, it is first translated into an arithmetic circuit using addition and multiplication gates. This circuit is then converted into a constraint system, such as the Rank-1 Constraint System (R1CS)~\cite{parno2016pinocchio}, which generalizes arithmetic circuit satisfiability. In R1CS, additions are represented by row vectors, for instance, $[0,1,0,1]$ for $(x_1 + w)$, while multiplications are encoded using element-wise products. Therefore, proving the original computation's correctness equals satisfying the R1CS. Efficient R1CS checking involves encoding it into polynomials. Schemes like \cite{parno2016pinocchio, groth16} use Quadratic Arithmetic Programs (QAP) to encode R1CS, derived through polynomial interpolation on R1CS instances. Other approaches employ univariate polynomials\cite{maller2019sonic, chiesa2020marlin} or multivariate polynomials~\cite{wahby2018hyrax,setty2020spartan} for encoding.

\begin{table}[t]
\centering
\scriptsize
\caption{The comparison between zkVC and prior verifiable DNN methods including SafetyNets~\cite{ghodsi2017safetynets}, Keuffer's~\cite{DBLP:conf/esorics/Keuffer}, vCNN~\cite{lee2020vcnn} , VeriML~\cite{DBLP:journals/tpds/veriml}, ZEN~\cite{cryptoeprint:2021/zen} , zkCNN~\cite{liu2021zkcnn}, zkML~\cite{kang2022scalingzkml} and pvCNN~\cite{weng2023pvcnn}.}
\begin{tblr}{
    colspec = {c |c c c c c c c cc},
    row{1-2} = {font=\bfseries},
    row{2-Z} = {rowsep=1pt},
    colsep = 2.5pt,
    }
\hline
\SetCell[r=2]{c}\textbf{Schemes} &\SetCell[r=2]{c}\textbf{zk.} & Non- & Const. & No Trusted & Trans- & Efficient  &zk-ML \\
&  & Inter. & Proof  &Setup & formers & MatMult &Codesign \\
\hline
SatetyNets&$\xmark$ &$\xmark$ & $\xmark$  &$\cmark$ &$\xmark$  &$\xmark$ &$\xmark$\\
zkCNN &$\cmark$      &$\xmark$ & $\xmark$  &$\cmark$ &$\xmark$ &$\xmark$ &$\xmark$\\ 
Keuffer's&$\cmark$  &$\cmark$ & $\cmark$  &$\xmark$ &$\xmark$  &$\xmark$ &$\xmark$\\ 
vCNN &$\cmark$      &$\cmark$ & $\cmark$  &$\xmark$ &$\xmark$  &$\xmark$ &$\xmark$\\
VeriML&$\cmark$     &$\cmark$ & $\cmark$  &$\xmark$ &$\xmark$  &$\xmark$ &$\xmark$\\ 
ZEN &$\cmark$       &$\cmark$ & $\cmark$  &$\xmark$ &$\xmark$ &$\xmark$ &$\xmark$\\ 
zkML&$\cmark$     &$\cmark$ & $\xmark$  &$\xmark$ &$\xmark$ &$\xmark$ &$\xmark$\\ 
pvCNN&$\cmark$      &$\cmark$ & $\cmark$  &$\xmark$ &$\xmark$ &$\xmark$ &$\xmark$\\
\hline
\textbf{zkVC} &$\cmark$ & $\cmark$  &$\cmark$&$\cmark$ &$\cmark$ &$\cmark$ &$\cmark$\\
\hline
\end{tblr}
\label{tab:prior}
 \vspace{-0.1in}
\end{table}

\begin{figure}
    \captionsetup{skip=1.2pt}
    \centering
    \includegraphics[width=0.8\linewidth]{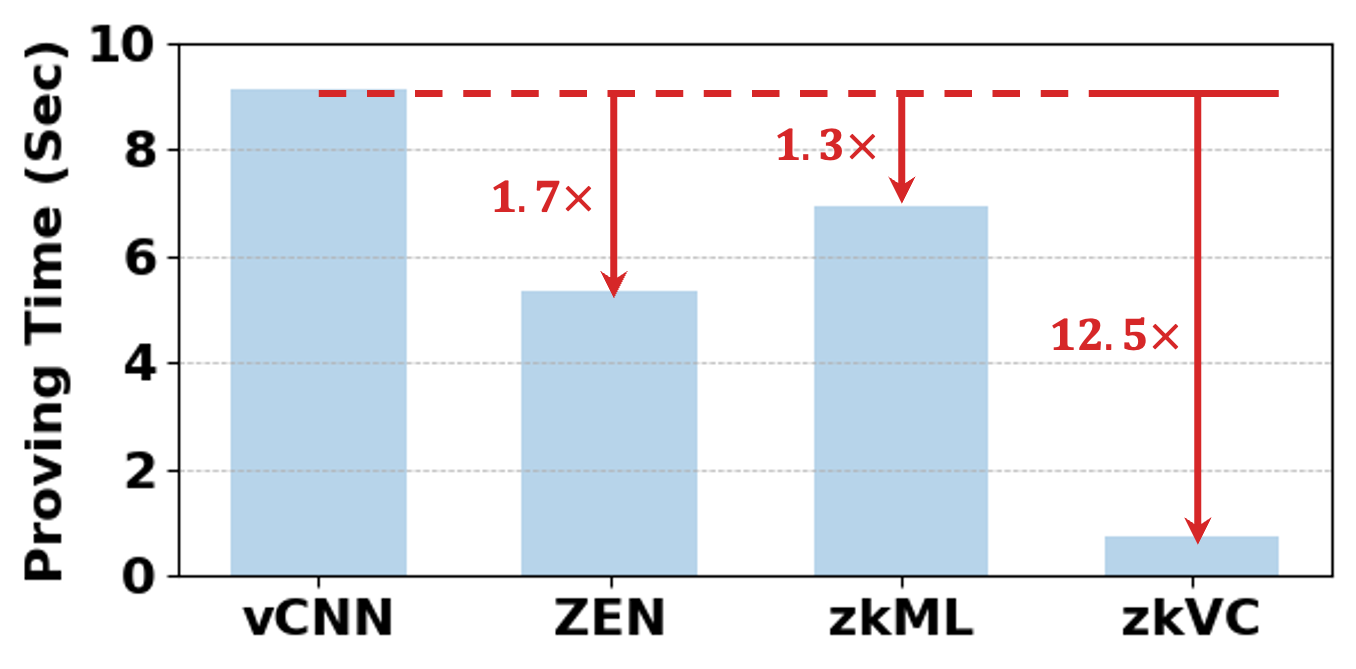}
     % \vspace{-0.1in}
    \caption{Proving Time Comparison for Matrix Multiplication with Prior Work.}
    \label{fig:motivation}
     \vspace{-0.3in}
\end{figure}

\begin{figure*}
% \vspace{-0.4in}
    \centering    \includegraphics[width=0.85\linewidth]{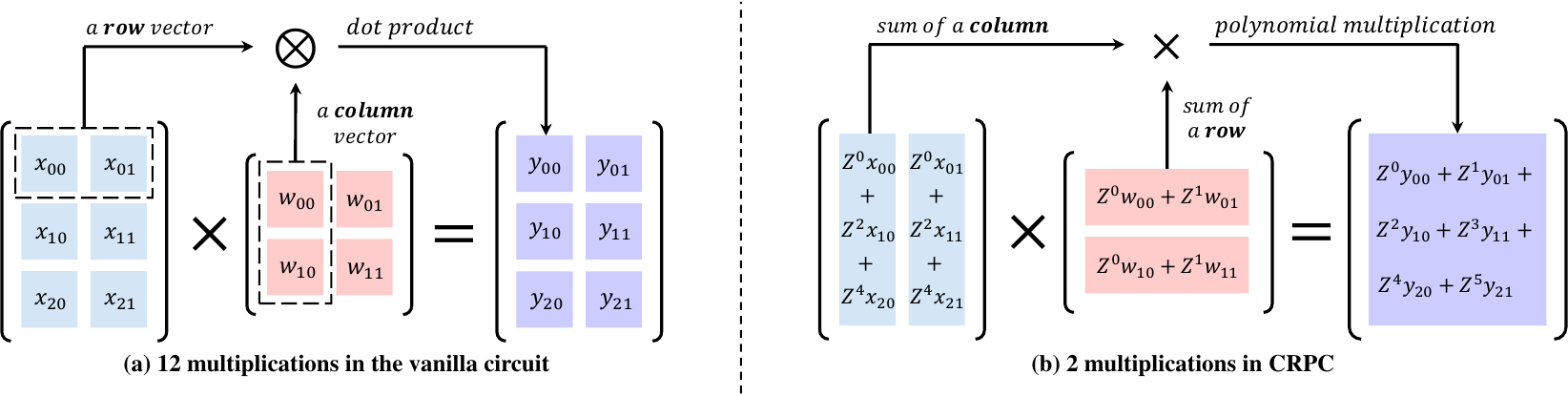}
    %\vspace{-0.2in}
    \caption{Comparison of (a) basic constraint circuits with $12$ multiplications and our (b) CRPC with  $2$ multiplications  by transforming original matrix multiplication into polynomial multiplications of an intermediate variable $Z$.}
    \label{fig:pom}
 \vspace{-0.2in}
\end{figure*}

The verification of R1CS or its polynomial equations requires specific cryptographic tools like groth16~\cite{groth16} or Spartan~\cite{setty2020spartan}. groth16, a widely-used ZKP construction, employs probabilistically checkable proofs (PCPs) and elliptic curve pairings for QAP verification. Spartan, on the other hand, uses Interactive Oracle Proof (IOP)\cite{ben2016iop} and polynomial commitment schemes\cite{bunz2020commit} for multivariate polynomial checks, notably without requiring a trusted setup. A ZKP construction consists of three Probabilistic Polynomial Time (PPT) algorithms: Gen, Prove, and Verify. Gen creates public parameters from security parameters. Prove uses these parameters, public and private inputs, to produce a proof. Verify, using the public parameters, input, and proof, decides its validity. ZKP ensures proof completeness and soundness, meaning correct computations pass verification, and it is computationally hard for a dishonest prover to validate incorrect computations.

% Verifying non-linear computations in ZKP requires either bit decomposition or approximation. For instance, the ReLU function can be approximated as a quadratic function compatible with arithmetic circuits~\cite{ghodsi2017safetynets}, or proven directly via bit decomposition~\cite{parno2016pinocchio, liu2021zkcnn}. This involves verifying that each bit is binary, the recomposed bits match the original input, and the output equals the sign bit's product with the recomposed unsigned bits. Bit decomposition also supports operations like fixed-point division and max pooling~\cite{liu2021zkcnn}. Linear and non-linear layers can be individually proven in sub-circuits and combined into a single proof using methods like commit-and-prove~\cite{campanelli2019legosnark}, as seen in approaches by ~\cite{DBLP:conf/esorics/Keuffer,lee2020vcnn} and ~\cite{weng2023pvcnn}, which encode computations in separate sub-circuits. 

%Apart from R1CS, there are also other constraint systems such as Plonkish~\cite{DBLP:journals/iacr/plonk} and Arithmetic Intermediate Representation (AIR)~\cite{ben2019scalable-air}. Various ZKP schemes are built with different cryptographic tools. Despite the variety of ZKP constructions, the proving complexity is mainly determined by the scale of the constraint system and number of constraints in the constraint system. 

\noindent \textbf{Homomorphic Encryption for Private Computation.} Homomorphic encryption (HE) enables computations directly on encrypted data, removing the need for decryption and thus supporting privacy-preserving outsourcing of computations~\cite{lou2019she,zheng2023primer,lou2021safenet,lou2021hemet}. However, HE alone does not inherently offer computational verifiability as ZKP do~\cite{lou2023vfhe, santriaji2024dataseal}. Conversely, while ZKP ensures computational integrity, it lacks the data privacy guarantees provided by HE~\cite{kumar2025tfhe, NEURIPS2024_Zhang, zhang2025cipherprune,deng2024trinity}. %Moreover, it is computationally intensive and currently scales only to small datasets such as MNIST and CIFAR-10~\cite{lou2021safenet,armknecht2015guide}, making it impractical for more complex or larger-scale tasks.

\noindent \textbf{Comparison with Related Work.} Table~\ref{tab:prior} shows the comparison of our zkVC and related works. SafetyNets~\cite{ghodsi2017safetynets} lacks zero-knowledge properties, leaving model weight privacy unprotected. Only SafetyNets and zkCNN~\cite{liu2021zkcnn} are interactive, necessitating ongoing communication between prover and verifier. Interactive ZKPs offer quicker proving times but require ongoing exchanges between the prover and verifier. While zero-knowledge polynomial commitment~\cite{wahby2018commit} and the Fiat-Shamir heuristic~\cite{fiat1986} could theoretically make these non-interactive, their security and efficiency impacts are unclear. The need for constant connectivity in interactive setups is a limitation, particularly for clients with limited hardware and power resources~\cite{cryptoeprint:2021/zen}. {Table~\ref{tab:prior}'s third column indicates that in SafetyNets, zkCNN, and zkML's~\cite{kang2022scalingzkml}, proof size grows logarithmically with model size, increasing verifier workload. Other schemes maintain a constant proof size.} Non-interactive schemes generally require a trusted setup for public parameter generation, but interactive ones like SafetyNets and zkCNN do not. zkVC, using transparent zk-SNARKs such as Spartan, also bypasses the need for a trusted setup. Previous research concentrated solely on CNNs, while zkVC focuses on general matrix multiplication, introducing efficient modules for proving matrix multiplication.

\noindent \textbf{Motivation.} Matrix multiplication is essential in many applications but challenging to prove using ZKP. Figure \ref{fig:motivation} shows that in prior vCNN~\cite{lee2020vcnn}, proving a small matrix multiplication of dimension $[49, 64]\times[64,128]$ takes as long as 9 seconds. Despite ZEN~\cite{cryptoeprint:2021/zen} introducing advanced quantization and zkML's~\cite{kang2022scalingzkml} using a more efficient ZKP method, their improvements in matrix multiplication speed are still limited. This motivates us to design efficient ZKP modules for matrix multiplication. The proposed zkVC achieves a 12.5$\times$ reduction in proof time compared to the previous vCNN.

% The complex non-linear functions, fundamental to the transformer's self-attention, cannot be directly supported into arithmetic circuits for ZKP. To solve this problem, we introduce a ZKP-compatible approximation for non-linear functions, including $SoftMax$. Figure \ref{fig:motivation}(b), evaluated on ViT-small on CIFAR-10, demonstrates that this $SoftMax$ approximation minimally impacts accuracy while being integrable with ZKP. To lower computational demands, we aim to find a ZKP-friendly token mixer with fewer number of non-linear $SoftMax$ functions. Attempts to replace all SoftMax self-attention with linear alternatives like average pooling, as suggested by~\cite{yu2022metaformer, lee2021fnetmixer}, resulted in notable accuracy losses, over 10\%. Combining approximated $SoftMax$ attention with pooling-based attention modules could lead to better accuracy or efficiency. However, manually searching for the hybrid combination led to suboptimal and unstable latency and accuracy. This led us to develop a ZK-friendly transformer planner for automated hybrid approximation search, balancing latency and accuracy optimally.

\section{zkVC Design}

\subsection{Constraint-reduced Polynomial Circuits (CRPC)}
\label{sec:crpc}
% To improve matrix multiplication proving efficiency in ZKP, we introduce CRPC. Figure \ref{fig:pom} (a) shows that naive matrix multiplication encoding uses many multiplication gates and is inefficient due to distinct constraints for each multiplication. In this method, each $y_{ij}$ is a dot product of $X$'s row and $W$'s column, requiring 12 constraints in the basic circuit. The high constraint number slows ZKP proving; CRPC is proposed to reduce this number. Our approach is based on the fact that each $y_{ij}$ in $\{x_{ik} \cdot w_{kj}\}_{k=0}^{n-1}$ comes from multiplications between the $k_{th}$ column of $X$ and the $k_{th}$ row of $W$. In CRCP, as shown in Figure \ref{fig:pom} (b), we convert columns of $X$ and rows of $W$ into polynomials of a random variable $Z$. For example, $X$'s first column becomes $X_0(Z) = Z^0x_{00} + Z^2x_{10} + Z^4x_{20}$. Correspondingly, matrix $Y$ transforms to $Y(Z) = Z^0y_{00}+Z^1y_{01}+...+Z^5y_{21}$, reducing constraints from 12 to 2 and keeping each $y_{ij}$'s accuracy. CRPC's proof is detailed in the appendix.

Matrix multiplication plays a foundational role in various computational tasks, but its efficient representation in QAP remains challenging. As in Figure \ref{fig:pom} (a), in vanilla QAP, every individual multiplication requires a distinct constraint. As every $y_{ij}$ is a dot product between a row vector of $X$ and a column vector of $W$, we have $y_{00} = x_{00} \cdot w_{00} + x_{01} \cdot w_{10}$, $...$ and $y_{21} = x_{20} \cdot w_{01} + x_{21} \cdot w_{11}$,
% \begin{equation*}
% \begin{aligned}
% y_{00} &= x_{00} \cdot w_{00} + x_{01} \cdot w_{10}\hfill\\
% \dots \hfill\\
% y_{21} &= x_{20} \cdot w_{01} + x_{21} \cdot w_{11}\hfill
% \end{aligned}
% \end{equation*}
where $12$ multiplications result in $12$ constraints in the QAP. Consider the matrix multiplication $Y = X \times W$ where $X \in  \mathbb{R} ^ {a \times n}$, $W \in  \mathbb{R} ^ {n \times b}$ and $Y \in  \mathbb{R} ^ {a \times b}$. Our insight is that the products in $\{x_{ik} \cdot w_{kj}\}_{k=0}^{n-1}$ for $y_{ij}$ can be encoded in polynomial multiplications. One intuitive transformation is:
\begin{equation*}
\begin{aligned}
(y_{00}+y_{01}+y_{10}+y_{11}+y_{20}+y_{21})\\
= (x_{00} + x_{10} + x_{20} ) \cdot (w_{00} + w_{01}) \\
+\ (x_{01} + x_{11} + x_{21}) \cdot (w_{10} + w_{11})
\end{aligned}
\end{equation*}
When matrix multiplication is satisfied, the above equation is also satisfied. However, the converse is not guaranteed. It is possible that the sum of all $y_{ij}$ is correct while individual $y_{ij}$ is not. Essentially, this transformation ensures completeness but compromises the soundness required by zk-SNARKs.

\begin{figure*}[t]
% \vspace{-0.4in}
    \centering
\includegraphics[width=0.85\linewidth]{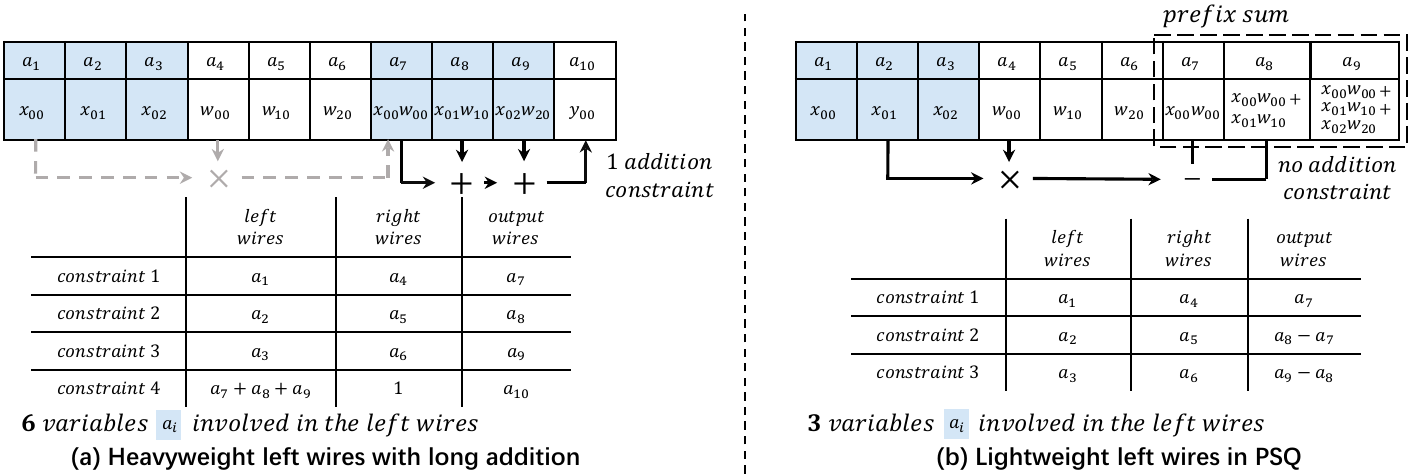}
    %\vspace{-0.1in}
    \caption{ Comparison: (a) traditional long addition with 6 left-wire variables vs. (b) our PSQ using only 3 variables.}
    %Comparing (a) traditional long addition involving 6 variables in the left wires, and (b) our PSQ which involves only 3 variables. }
    \label{fig:prefix}
 \vspace{-0.2in}
\end{figure*}

As prior work's approach~\cite{lee2020vcnn} suggests, it is possible to encode the matrix multiplication in one single polynomial multiplication, if the coefficients are properly arranged. Another possible transformation is:
\begin{equation*}
\begin{aligned}
(Z^1y_{00}+Z^3y_{01}+Z^5y_{10}+Z^7y_{11}+Z^9y_{20}+Z^{11}y_{21})\\
\ne (Z^1x_{00} + Z^0x_{01} + Z^5x_{10} + Z^4x_{11} + Z^9x_{20} + Z^8x_{21}) \\ 
\cdot\ (Z^0w_{00} + Z^2w_{01} + Z^1w_{10} + Z^3w_{11})
\end{aligned}
\end{equation*}
where $y_{00} = x_{00} \cdot w_{00} + x_{01} \cdot w_{10}$ is incorporated within $Z^1y_{00} = Z^1(x_{00} \cdot w_{00} + x_{01} \cdot w_{10})$. However, this transformation is not strictly equivalent. Firstly, the equation itself is not satisfied. The polynomial multiplication results in a lot of superfluous terms such as $Z^0x_{01}w_{01}$, which does not belong to any $y_{ij}$. Secondly, to equate the two sides, all these dummy terms need to be included, which leads to increased number of variables and makes the circuit hard to prove.

While the above two transformations can reduce the number of multiplications, they cannot ensure the integrity of matrix multiplication. We propose CRPC to address this issue. Our insight is that the products in $\{x_{ik} \cdot w_{kj}\}_{k=0}^{n-1}$ for $y_{ij}$ result only from the multiplications between elements in the $k_{th}$ column in $X$ and elements in $k_{th}$ row in $W$. As shown in Figure \ref{fig:pom} (b), we transform each column of $X$ and each row of $W$ into polynomials of a random intermediate variable Z, for example, the first column of $X$ is $X_0(Z) = Z^0x_{00} + Z^2x_{10} + Z^4x_{20}$. The matrix $Y$ is converted accordingly. We have:
% $Y(Z) = \sum_{k=0}^{n-1} (X_k(Z)\cdot W_k(Z))$
\begin{equation*}
\begin{aligned}
(Z^0y_{00}+Z^1y_{01}+Z^2y_{10}+Z^3y_{11}+Z^4y_{20}+Z^5y_{21})\\
= (Z^0x_{00} + Z^2x_{10} + Z^4x_{20} ) \cdot (Z^0w_{00} + Z^1w_{01}) \\
+\ (Z^0x_{01} + Z^2x_{11} + Z^4x_{21}) \cdot (Z^0w_{10} + Z^1w_{11})
\end{aligned}
\end{equation*}
where only 2 multiplications are needed. We generalize CRPC's transformation for matrix multiplication \(Y^{a \times b} = X^{a \times n} \times W^{n \times b}\) as:
\begin{equation*}
\begin{aligned}
\sum_{j=0}^{b-1} \sum_{i=0}^{a-1} Z^{ib+j}y_{ij} = \sum_{k=0}^{n-1} \left ( \sum_{i=0}^{a-1}Z^{ib}x_{ik}  \right ) \left ( \sum_{j=0}^{b-1}Z^{j}w_{kj}  \right )
\end{aligned}
\end{equation*}
where only  \(n\) constraints are needed, where as  \(a \cdot b \cdot n\) constraints are needed in vanilla circuits.

Polynomial multiplication offers significant advantages in computational efficiency, especially when representing matrix multiplication in ZKP protocols. Our proposed CRPC reduces the constraint complexity from \(O(n^3)\) to \(O(n)\) for matrix multiplication. The proving efficiency of zk-SNARKs systems is thereby improved by a significant margin. %Computing the $H$ query can be up to $18000\times$ faster in setup time. Computing the quotient polynomial and H query can be $40\times$ and $8000\times$ faster in prove time. 
The overall proving time of ZKP-based matrix multiplication of different sizes in Transformer layers can be $7\sim9\times$ faster.

% \qnote{consider if specific matrix multiplication size is needed for this $9\times$ improvement}.

\begin{figure*}
    \centering
    \includegraphics[width=1\linewidth]{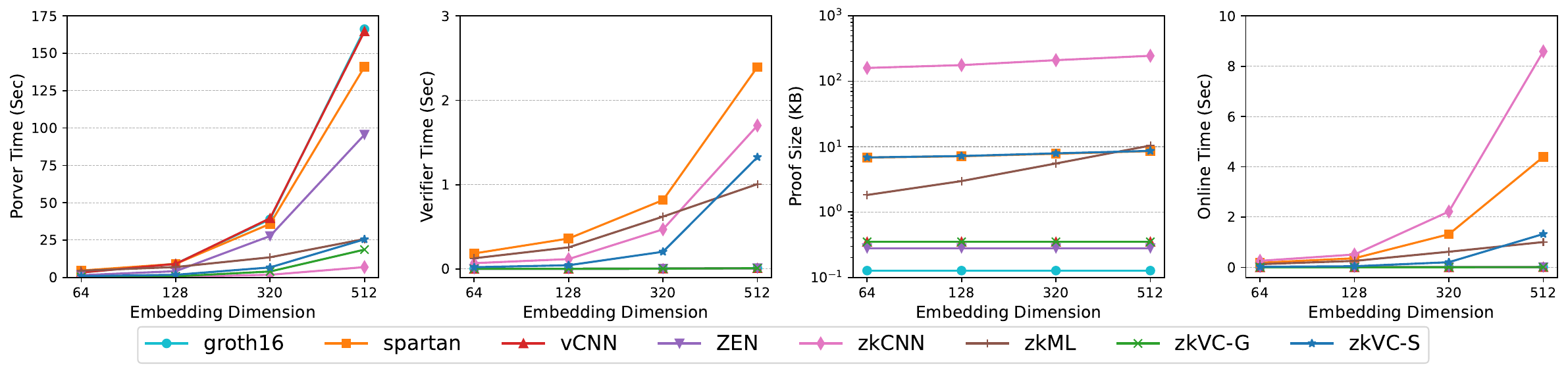}
    %\vspace{-0.25in}
    \caption{A comparison of zkVC and prior works on matrix multiplication. zkVC leads in proving time of all non-interactive methods, close to interactive zkCNN, and excels in verification time, proof size, as well as online verification duration. }
    \label{fig:matmult}
    \vspace{-0.2in}
\end{figure*}

\subsection{Prefix-Sum Query (PSQ)}
\label{sec:psq}

The CRPC method notably decreases the constraint count needed for matrix multiplication representation. Additionally, it is observed that the number of left wires influences proving performance. We propose PSQ to reduce the number of left wires.  Our insight is that, although an arbitrary number of additions can be encapsulated within a single constraint, a prolonged sequence of additions can result in a considerable computational overhead, leading to a large number of left wires number, as is shown in Figure~\ref{fig:prefix}(a). 
Consider a dot product in matrix multiplication, represented as $y_{00} = x_{00} \cdot w_{00} + x_{01} \cdot w_{10} + x_{02} \cdot w_{20}$. This calculation requires $4$ constraints. The initial three constraints calculate the intermediate products: $x_{00}w_{00}$, $x_{01}w_{10}$, and $x_{02}w_{20}$, which are assigned to variables $a_7$, $a_8$, and $a_9$. To achieve the final result $y_{00}$, one more addition is required, incorporating the three intermediate variables $a_7$, $a_8$, and $a_9$ in the left wires. Consequently, this approach uses $6$ left variables/wires. Heavy left wires in large matrices can be computationally demanding. Then, we introduce PSQ to avoid extended additions in matrix multiplication, as illustrated in \ref{fig:prefix} (b). Instead of holding actual intermediate product values in $a_7$, $a_8$, and $a_9$, we record their prefix sums.  Specifically, 
$a_7 = x_{00}w_{00}, \quad a_8 = a_7 + x_{01}w_{10}, \quad a_9 = a_8 + x_{02}w_{20}.$

Here, the final result $y_{00}$ is directly available in $a_9$, eliminating the need for an additional constraint for the long addition. PSQ reduces the left wire variables to only $3$, making it a more efficient approach for both verification and proving stages.

The proposed PSQ effectively reduces the complexity. Consider the general matrix multiplication \(Y^{a \times b} = X^{a \times n} \times W^{n \times b}\). There are $(a\cdot n + a\cdot b \cdot n)$ variables involved, where the $(a\cdot b \cdot n)$ intermediate products make proving rather complex. With PSQ, the proving is only associated with the $(a\cdot n)$ variables, and the complexity is reduced from $O(n^3)$ to $O(n^2)$. PSQ effectively contributes to a lightweight R1CS, and the computation of the R1CS thus becomes significantly more efficient. Specifically, the cost of computing the R1CS is reduced by approximately $70\%$ during the proving phase. By building on the foundation of CRPC, PSQ further reduces the proving cost for matrix multiplication by $30\%$. This results in a total speedup of $12\times$ for the proving time.

\subsection{Nonlinear-Function Approximation}
\label{subs:non-linear}
CRPC and PSQ significantly improves the efficiency of proving matrix multiplication with ZKP. To demonstrate the efficiency of our zkVC, we apply it to verifiable Transformer inference. To verify the correctness of computation in transformers, we need to verify the non-linear functions like $SoftMax$, and GeLU with ZKP. However, ZKP can not directly support these non-arithmetic functions.  We design accurate arithmetic approximations for these complex non-arithmetic functions. Given a vector $x\in\mathbb{R}^d$, the $SoftMax$ function is defined as $SoftMax_i(x) = e^{x_i} / \sum_{j\in[d]}e^{x_j}$. The main challenge is to accurately express the exponential function in ZKP constraints. Although the exponential function can not be directly represented by addition and multiplication, it is possible to closely approximate the exponential function on negative inputs. Based on this idea, we show how the SoftMax function is verified in our design as follows. 

\noindent\textbf{Computing $x_{max}$.} We first normalize the input vector $x$ by $(x-x_{max})$, where $x_{max}$ is the max element in $x$. The subtraction can easily be encoded in ZKP circuits. To verify the max computation, we check two constraints: (1) $x_{max} \geq x_j $ for all $j \in [d]$ and (2) $\prod_{j}(x_{max}-x_j)=0$. The first constraint ensures $x_{max}$ is greater than all other values in $x$ and the second constraint ensures $x_{max}$ is indeed one of the values from $x$. Since ZKP supports comparison operations by bit-decomposition~\cite{liu2021zkcnn}, these constraints are compatible with ZKP.

\noindent\textbf{Approximating $e^x$ on negative inputs.} After normalizing $x$ by $(x-x_{max})$, all elements in the resulting vector are negative. 
It is evident that computing $SoftMax(x - x_{max})$ is equivalent to computing $SoftMax(x)$. Thus, the SoftMax function can be computed as $SoftMax_i(x) = e^{x_i - x_{max}} / \sum_{j \in [d]} e^{x_j - x_{max}}$, where only the exponential function on negative inputs is needed to compute. We approximate $e^x$ on negative inputs using the Taylor series:
\begin{equation*}
\begin{aligned}
e^x \approx
\begin{cases} 0,  & \text{if }x < T \\
(1+x/2^n)^{2^n}, & \text{if }x \in [T,0].
\end{cases}
\end{aligned}
\end{equation*}
where T is the pre-defined threshold deciding the clipping branch. We apply a two-bit decomposition to compare $x$ with $T$ and compute the division by $2^n$. The power-to-$2^n$ computation is essentially a series of multiplications that can be easily encoded in ZKP circuits.

Putting together, we verify the SoftMax function in ZKP via a close approximation using three sets of bit decomposition and two sets of multiplication. %Such approximation has been show to have only negligible average error within $2^{-10}$~\cite{lu2023bumblebee}, and to the best of our knowledge, we are the first to explore verifying the SoftMax function with ZKP. 
The GELU activation function is used in NLP transformers such as BERT. The GELU function is defined as $\mathsf{GELU}(x)=0.5x\left( 1+\mathsf{Tanh}[\sqrt{2/\pi}(x+0.044715x^3)] \right)$. Similar to the exponential function in $SoftMax$, the $\mathsf{Tanh}$ function can not be directly represented by addition and multiplication gates. We use polynomial approximation to represent the GELU function efficiently. Specifically, we use $\mathsf{GELU}(x) \approx x^{2}/8 + x/4+ 1/2$.%~\cite{li2022mpcformer}.

\section{Experimental Methodology}

\noindent\textbf{Models and Datasets.} In our experiments, we explored both computer vision and language transformers. For computer vision, we tested three vision transformer architectures on different datasets, adapting settings from~\cite{hassani2021samllvit, zeng2022mpcvit} for small datasets. On CIFAR-10, the ViT configuration was 7 layers, 4 heads, a hidden dimension of 256, and a patch size of 4. For Tiny-ImageNet, we used 9 layers, 12 heads, a hidden dimension of 192, and a patch size of 4. On ImageNet, a hierarchical architecture from~\cite{liu2021swin,yu2022metaformer} was implemented with 12 layers and 4 stages, with embedding dimensions of 64, 128, 320, 512. For NLP, we chose a BERT model with 4 layers, 4 heads, and an embedding dimension of 256. This model was fine-tuned and assessed on GLUE benchmarks~\cite{wang2018glue}, including MNLI, QNLI, SST-2, and MRPC tasks.

\noindent\textbf{Implementation Details.}
%We implemented zkVC's efficient modules, CRPC and PSQ in C++ and Rust. 
In our experiments, we utilized groth16 from libsnark~\cite{libsnark} and Spartan~\cite{setty2020spartan} for the ZKP backend, referred to as zkVC-G and zkVC-S, respectively. These cryptographic tests were conducted on AMD Ryzen Threadripper PRO 3955WX 16-Core CPU systems with 128GB RAM, running Ubuntu 22.04.1. Transformer model experiments were performed on NVIDIA GeForce RTX 3090 GPUs. For Transformer architectures, zkVC was built on ViT~\cite{dosovitskiy2020vit} and MetaFormer's frameworks~\cite{yu2022metaformer}, incorporating efficient token mixers like scaling attention modules~\cite{shen2020linearE, wang2018scaleattn}, linear transformation modules~\cite{lee2021fnetmixer}, average pooling, and differentiable NAS~\cite{liu2018darts}. We employed the quantization technique from~\cite{Wang_2022/quantize} for converting model parameters to integer formats. This marks the first instance of verifiable Transformers being tested on the ImageNet dataset~\cite{deng2009imagenet}.

\section{Results}

\subsection{Micro-benchmarks}
\label{sec:zkmodule}

\noindent \textbf{Matrix Multiplication Benchmark.}
Figure \ref{fig:matmult} shows a comparison of zkVC with previous works on matrix multiplication benchmarks. The dimensions of the matrices are set according to the embedding layers in ViT~\cite{dosovitskiy2020vit}, i.e., $[\#tokens, dim_1]\times[dim_1, dim_2]$. The number of tokens are set to $49$ for simplicity. For example, when the embedding dimension is $128$, the dimension of matrix multiplication is $[49, 64] \times [64, 128]$. For non-interactive schemes, zkVC-G is based on groth16~\cite{groth16}, and zkVC-S on Spartan~\cite{setty2020spartan}. vCNN~\cite{lee2020vcnn} and ZEN~\cite{cryptoeprint:2021/zen} also use groth16, while Kang's~\cite{kang2022scalingzkml} employs halo2~\cite{bowe2019halo2}. The interactive scheme included is zkCNN~\cite{liu2021zkcnn}, with vanilla groth16 and Spartan serving as baselines. zkVC significantly enhances proving time, achieving 5 to 12 times faster performance than the groth16 and Spartan baselines. Among non-interactive options, zkVC-G stands out for its proving efficiency. Though zkCNN is about twice as fast in proving compared to zkVC-G, it requires interaction and suffers from slower verification and larger proof sizes. Specifically, zkCNN's verification is up to 200 times slower and its proofs are 1 to 2 orders of magnitude larger than those of zkVC. The online time means the time that the client and server need to maintain online during the proving. zkCNN also demands additional verifier online time due to its interactive nature.

\noindent \textbf{Technical Ablation Study.}
In Table \ref{tab:ablation}, we present the latency results from a matrix multiplication microbenchmark to highlight the effectiveness of the proposed CRPC and PSQ in transformer patch embedding layers. CRPC significantly reduces proving time, boosting the groth16 backend by approximately 9$\times$ and the Spartan backend by around 5$\times$. While groth16 maintains a consistent verification time, CRPC cuts Spartan's verification time by about 4$\times$. The application of PSQ further speeds up groth16, achieving up to a 12$\times$ faster rate. Directly applying PSQ to Groth16 can also improve proving time, but the enhancement is more significant when PSQ is coupled with CRCP. However, PSQ's impact on Spartan's proving time is minimal, this is because CRPC already reduces the number of constraints for both backends, while PSQ primarily simplifies specific complex queries for groth16.

% \textbf{End-to-end performance.} Matrix multiplication serves as a fundamental component across diverse applications. We extend this understanding by showcasing the adaptability of the proposed CRP-QAP and PSQ, considering varying input dimensions of model architecture, as illustrated in Table \ref{t:cnn}. zkVC, accommodating various input resolutions, uniformly gains from the newly proposed efficient ZKP module, experiencing a substantial acceleration in the range of $12\times$ to $20\times$.

\begin{table}[t]
\centering
\scriptsize
\caption{Ablation study on matrix multiplication microbenchmark.}
\begin{tblr}{
    colspec = {cc|cc|cc},
    row{1} = {font=\bfseries},
    row{2-Z} = {rowsep=1pt},
    colsep = 4.5pt,
    }
\hline
\SetCell[r=2]{c}\textbf{CRPC} &\SetCell[r=2]{c}\textbf{PSQ}  &\SetCell[c=2]{c}\textbf{groth16} &&\SetCell[c=2]{c}\textbf{Spartan} \\ 
&& \textbf{Prove(s)} &\textbf{Verify(s)} & \textbf{Prove(s)} &\textbf{Verify(s)}\\
\hline
\xmark &\xmark &9.12 & 0.002 & 9.04 & 0.36\\
\xmark &\cmark &8.69 &0.002 &8.95 &0.32 \\
\cmark &\xmark &1.01 & 0.002 & 1.79 & 0.08\\
\cmark &\cmark &0.73 & 0.002 & 1.75 & 0.05\\
\hline
\end{tblr}
\vspace{-0.1in}
\label{tab:ablation}
\end{table}

\begin{table}[t]
\centering
\scriptsize
\caption{Comparison of various token mixers with our zkVC on ViT Models.  SoftApprox. includes approximated SoftMax, Softfree-S (scaling), Softfree-P (pooling), with $\mathcal{P}_G$ for Groth16 and $\mathcal{P}_S$ for Spartan. }
\begin{tblr}{
    colspec = {cc|c|cc},
    row{1} = {font=\bfseries},
    row{2-Z} = {rowsep=1pt},
    colsep = 4pt,
    % row{5} = {bg=LightBlue},
    }
\hline
\textbf{Dataset} &\textbf{Model} 
&Top1(\%) &\textbf{$\mathcal{P}_G$(s)} &\textbf{$\mathcal{P}_S$(s)}\\
\hline
\SetCell[r=4]{c}\textbf{Cifar-10}
&SoftApprox. &93.5&725.2&1006.2\\
&SoftFree-S &88.3&568.4&742.8\\
&SoftFree-P &75.1&262.7&300.6\\
&\hline zkVC &91.6&458.6&591\\
\hline
\SetCell[r=4]{c}\textbf{Tiny ImageNet}
&SoftApprox. &60.5&1609.6&2197.4\\
&SoftFree-S  &51.4&1004.9&1348.8\\
&SoftFree-P &42.7&443.7&503.6\\
&\hline zkVC &55.8&879.3&1161.4\\
\hline
\SetCell[r=4]{c}\textbf{ImageNet}
&SoftApprox. &81&10700&12857.7\\
&SoftFree-S  &78.5&4521.3&5812.7\\
&SoftFree-P &77.2&2904&3667.8\\
&\hline zkVC &80.3&3457.1&4417.1\\
\hline
\end{tblr}
\vspace{-0.25in}
\label{tab:end_cv}
\end{table}

\begin{table}[t]
\centering
\scriptsize
\caption{Comparison of various token mixers with our zkVC on NLP Models.  SoftFree-L denotes a model using linear transformation for token mixing.}
\begin{tblr}{
    colspec = {c|cccc|cc},
    row{1} = {font=\bfseries},
    row{2-Z} = {rowsep=1pt},
    colsep = 5.5pt,
    }
\hline
\SetCell[r=2]{c}\textbf{Model} &\SetCell[c=4]{c}\textbf{Acc. on Tasks(\%)} &&&& \SetCell[r=2]{c}\textbf{$\mathcal{P}_G$(s)} &\SetCell[r=2]{c}\textbf{$\mathcal{P}_S$(s)} \\
& \textbf{MNLI} & \textbf{QNLI} & \textbf{SST-2} & \textbf{MRPC}
\\
\hline
SoftApprox. &74.5 &83.9 &85.8 &71.2 &1299.5 &1793.3\\
SoftFree-S &72.7 &81.1 &85.2 &70.4 &917.1 &1201.4\\
SoftFree-L &67.3 &75.3 &84.5 &68.7 &680.8 &782.0\\
\hline
zkVC &70.8 &80.2 &84.7& 69.3& 798.9 &992.2\\
\hline
\end{tblr}
\vspace{-0.2in}
\label{tab:end_nlp}
\end{table}

\subsection{End-to-end Performance}
\label{sec:planner}
%To showcase the effectiveness of our planner, we illustrate the trade-off between the accuracy and latency of various transformer architectures on both computer vision and NLP tasks. 

\noindent\textbf{Vision.}
Table \ref{tab:end_cv} illustrates zkVC's accuracy-latency balance on three popular vision datasets. For smaller datasets like Cifar-10 and Tiny ImageNet, using average pooling instead of SoftMax self-attention notably reduces accuracy. Although scaling attention models are more accurate, their efficiency gain in ZKP proving is minimal due to low-resolution images (e.g., $32\times32$ in Cifar-10) and fewer transformer input tokens (e.g., 64 tokens for 4-size patches). Average pooling alone struggles with low-res images, and scaling attention alone shows limited efficiency for short token sequences. zkVC, blending $SoftMax$ self-attention with $SoftMax$-free options, achieves around 40\% faster proving on Cifar-10 with under 2\% accuracy loss, and about 50\% faster on Tiny ImageNet with less than 5\% accuracy loss.

High-resolution ImageNet images, with many input tokens (e.g., 3136 for a $224\times224$ image with patch size 4), strain SoftMax self-attention's quadratic complexity. Scaling attention's linear complexity better manages these long sequences. SoftMax-free methods, like scaling attention and average pooling, excel on larger datasets, speeding up computations by 60\% to 70\% with less than 5\% accuracy loss. zkVC achieves similar speedups with under 1\% accuracy loss, likely due to reintegrating SoftMax self-attention in later transformer layers with shorter token sequences.

\noindent\textbf{NLP.}
Table \ref{tab:end_nlp} shows zkVC's evaluation on NLP transformers like BERT. Linear transformation improves efficiency by ~50\% but can drop accuracy by up to 7\% on MNLI. Scaling attention increases proving efficiency by 30\%, with better, more stable performance across tasks. zkVC is about 15\% faster than scaling attention models while being around 3\% more accurate on average compared to linear transformation. The results indicate that replacing all \textit{SoftMax} attention with SoftMax-free alternatives isn't always ideal for accuracy and latency demands. zkVC, using our planner, combines a hybrid transformer architecture. It matches \textit{SoftMax}-centric models in accuracy and outperforms Linear Attention models in latency.

% The rationale behind this hybrid architecture is clarified when we assess transformer blocks with varying token mixers across each stage, as visualized in Figure \ref{fig:tokenstage}. \textit{SoftMax}-based attention proves notably taxing during the transformer's initial two stages, where pooling and linear attention offer more efficient proving. However, merely supplanting \textit{SoftMax} attention with pooling and linear attention might compromise accuracy. We discern that the penultimate stage plays a pivotal role in preserving the transformer's accuracy. In this $stage_3$, we speculate that the introduction of linear attention underpins the efficacy of this mixed model. During the final stage, both \textit{SoftMax}-based attention and linear attention impose similar proving burdens—reintegrating \textit{SoftMax}-based attention further ensures accuracy.

\section{Conclusion}
In this paper, we introduced \textit{zkVC}, an efficient zk-SNARK construction designed to optimize matrix multiplication verification. Traditional ZKP approaches for verifying matrix multiplication often require an excessive number of constraints, leading to high computational overhead. \textit{zkVC} addresses this challenge by leveraging CPRC to minimize constraints and PSQ to reduce variables, achieving a $12\times$ improvement in proof efficiency over prior methods. We further demonstrated \textit{zkVC}'s effectiveness in verifiable Transformer inference, verifying the integrity of ViT models efficiently. Given that matrix multiplication underpins a wide range of applications, we believe \textit{zkVC} offers a significant step forward in enhancing the computational integrity of these applications.

% a novel solution enhancing Transformer prediction integrity by verifying origins while maintaining confidentiality. It introduces efficient ZKP modules like CRPC and PSQ, which reduce computational demands and achieve a significant 12$\times$ improvement in proof latency. zkVC also proposes non-linear approximation and an autonomous planner balancing performance and accuracy, boosting proving efficiency by 40\% to 70\%. %This positions zkFormer as a key advancement in deep learning credibility, especially for sensitive applications, by robustly verifying model inferences' authenticity and accuracy.

\bibliographystyle{IEEEtran}
\bibliography{IEEEexample}

% \begin{thebibliography}{00}
% \bibitem{b1} G. Eason, B. Noble, and I. N. Sneddon, ``On certain integrals of Lipschitz-Hankel type involving products of Bessel functions,'' Phil. Trans. Roy. Soc. London, vol. A247, pp. 529--551, April 1955.
% \end{thebibliography}

% \appendix
% \input{contents/appendix}

\end{document}